\documentclass[twocolumn,
showpacs,preprintnumbers,amsmath,amssymb,floatfix,aps]{revtex4}
\usepackage{graphicx}% Include figure files
\usepackage{dcolumn}% Align table columns on decimal point
\usepackage{bm}% bold math

\newcommand{\be}{\begin{equation}}
\newcommand{\ee}{\end{equation}}
\newcommand{\ua}{\uparrow}
\newcommand{\da}{\downarrow}

\begin{document}

%\preprint{APS/123-QED}

\title{Spin polarizability of a trapped superfluid Fermi gas}

\author{A. Recati, I. Carusotto, C. Lobo and S. Stringari}
% \altaffiliation[Also at ]{Physics Department, XYZ University.}
\email{recati@science.unitn.it} \affiliation{ $^{a}$Dipartimento di
Fisica, Universit\`a di Trento and CNR-INFM BEC Center, I-38050
Povo, Trento, Italy \\}

\begin{abstract}
The polarization produced by the relative displacement of the potentials trapping  two spin species of a dilute Fermi gas with $N_\ua=N_\da$ is calculated at unitarity by assuming phase separation between the superfluid and a spin polarized phase at zero temperature. Due to the energy cost associated with pair breaking, the dipole magnetic polarizability  vanishes in the linear limit and exhibits important deviations from the ideal gas behaviour  even for displacements of the order of the size of the atomic cloud. The magnetic behaviour in the presence  of  different trapping frequencies for the two spin species is also discussed.

\end{abstract}
%\pacs{Valid PACS appear here}

\maketitle

It is well known that Fermi superfluids cannot be polarized by an external magnetic field unless the field exceeds a critical value. This effect is directly associated with the occurrence of a gap in the spin excitation spectrum and is the consequence of the existence of pairs. In a dilute gas the nature of these pairs depends crucially  on the value and the sign of the s-wave scattering length. For negative and small values of the scattering length the pairs  coincide with the  Cooper pairs of BCS superconductivity.  If the scattering length is instead positive and small, they can be identified with real molecules which, due to their bosonic nature, undergo Bose-Einstein condensation (BEC).  With the aid of Feshbach resonances it is now possible to experimentally control the transition between the BCS and BEC regimes and to investigate  the challenging unitary regime where the scattering length is much larger than the interparticle distance. The effect of spin polarization on these novel quantum phases has already been the object of experimental \cite{Rice, MIT1, MIT2} and theoretical \cite{theory,chevy,erich} investigations. In particular the polarization has been shown to give rise to a phase separation between superfluid and  non-superfluid components, although the detailed structure of the phase separation is still far from being completely understood. 

The purpose of this paper is to investigate directly the response of these systems to a spin-dependent external field, taking advantage of the fact  that the trapping geometry suggests a natural way to generate an effective, position dependent, ``magnetic'' field. This is achieved, for example, by an adiabatic separation of the external potentials confining the two spin species. If the gas is noninteracting the two spin clouds will move rigidly in opposite directions giving rise to a spin dipole moment per particle 
\begin{equation}
\label{D}
D(d)={1\over N}\int x (n_\uparrow-n_\downarrow)d{\bf r} \; ,
\end{equation} 
equal to one half of the distance separating the two wells. In (\ref{D}) $N$ is the total particle number and $d$ the displacement of the trapping potential. If instead the system is interacting and superfluid, it will exhibit a resistance to spin polarization. In particular the linear response will vanish. 

In order to calculate the effects of the local polarization induced by the displacement of the confining potentials we will make use of the phase diagram of uniform matter and of the corresponding equations of state which are available in some relevant regimes,  the equilibrium between the different  phases being obtained  by imposing that the corresponding pressures be equal. 
In the following we will mainly focus on the unitary regime where the equation of state takes a universal form. In the unpolarized phase the system is superfluid at zero temperature and its equation of state is
\begin{equation}
\label{mus}
\mu_s=\frac{\hbar^2}{2m}\xi(6\pi^2 n)^{2/3}
\end{equation}
where $n=n_\ua=n_\da$, exhibiting the same power law density dependence as in the noninteracting gas. The dimensionless parameter $\xi$ accounts for the role of interactions. Its value, evaluated with {\it ab initio} Monte Carlo simulations, is $\xi\simeq 0.44$ (see, e.g., \cite{astraxi}). In the presence of polarization  we will assume 
that the system exhibits phase separation between an unpolarized superfluid phase (s) governed by the equation of state (\ref{mus}) and a fully spin polarized phase (p) where interactions can be ignored and the equation of state is  given by the noninteracting expression 
\begin{equation}
\label{musp}
\mu_{p}=\frac{\hbar^2}{2m}(6\pi^2 n)^{2/3}\; ,
\end{equation}
with $n$ the density of the only species present in such a phase. This is the simplest assumption on resonance. The more complicated possibility of an intermediate non-superfluid phase where the two components coexist with different densities has been recently explored theoretically \cite{theory,erich} and experimentally \cite{MIT2}. 

\begin{figure}[ptb]
\begin{center}
\includegraphics[height=4.3cm] {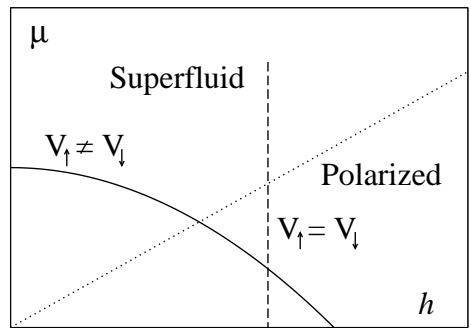}
\end{center}
\caption{Trajectories in the phase diagram for an unbalanced Fermi gas in a spin-independent potential (vertical dashed line) and for the dipole configuration discussed in the text (parabolic continouos line). The diagram exhibits a first order phase transition from the superfluid to a fully polarized phase along the diagonal dotted line.}
\label{fig:phases}
\end{figure}

In terms of the chemical potentials $\mu_s$ and $\mu_{p}$ the equilibrium between the pressures of the two phases  can be usefully written in the form 
\begin{equation}
\label{eqcond}
2\mu_s=(2\xi)^{3/5}\mu_{p}.
\end{equation}
It is useful to represent the equilibrium condition (\ref{eqcond}) in the $\mu-h$ phase diagram (see figure \ref{fig:phases}). Here $\mu= (\mu_\ua+\mu_\da)/2$ is the average value of the two chemical potentials while $h=(\mu_\ua-\mu_\da)/2$
is an effective magnetic field fixed by the difference between the two chemical potentials. From general thermodynamic considerations \cite{chevy} we find
\begin{equation}
\label{musud}
\mu_s = {1\over 2}(\mu_\ua+\mu_\da) \; ,
\end{equation}
so that for $h>0$, where $\mu_p=\mu_\ua$, the phase separation between the superfluid and the spin polarized phase is described by the straight line 
\be\mu=\frac{(2\xi)^{3/5}}{2-(2\xi)^{3/5}}h.\ee

The trapping introduces a position dependence in the chemical potentials of the two atomic spieces that can be evaluated  according to
the local density expressions
\begin{equation}
\label{muud}
\mu_{\ua\da}({\bf r})=\mu_{\ua\da}^0-V_{\ua\da}({\bf r})
\end{equation}
Result (\ref{muud}) allows us to calculate the $\bf r$-dependence of  the chemical potential of both the superfluid
and  spin-polarized phases.  The equilibrium condition (\ref{eqcond}) then becomes an equation characterizing the $\bf r$ dependence of the surface separating the two phases. The constant values $\mu_{\ua\da}^0$ are determined by imposing the proper normalization on the the spin-up and spin-down densities.

If the trapping potential is the same for the two species ($V_\ua=V_\da$) then
the effective magnetic field $h$, which differs from zero if $N_\ua\neq N_\da$,
is independent of position and, by varying $\bf r$, we are consequently exploring the $\mu-h$ phase space  at fixed  $h$, as shown by the vertical line in Fig. \ref{fig:phases}. 
This is the case of all the configurations  so far investigated in the literature. If instead, the two external potentials are different, the magnetic field $h$ will be position dependent and, by varying $\bf r$, we explore the phase diagram along the parabolic line shown in the same figure. We will discuss in particular the case of a dipole displacement of two harmonic traps  
%\be V_{\uparrow\downarrow}({\bf r})=\frac{1}{2}m\omega^2(y^2+z^2+(x\mp {1\over 2d^2). \ee
\be V_{\uparrow\downarrow}({\bf r})=\frac{1}{2}m\left(\omega_\perp^2 r_\perp^2+\omega_x^2(x\mp d)^2\right), \ee
with $r_\perp=\sqrt{y^2+z^2}$. 
Furthermore we will assume that the system to be globally unpolarized ($N_{\ua}=N_{\da}\equiv N/2$) so that, for symmetry reasons,
$\mu_\ua^0=\mu_\da^0\equiv \mu_0$ and the magnetic field is simply given by the $x$-dependent expression $h=m\omega^2_x x d$.

It is worth noticing that for any value of the displacement $d$  the effective potential $(V_\ua({\bf r})+V_\da({\bf r}))/2$ felt by the system  in the superfluid phase has  a minimum at $x=0$ giving rise to a superfluid density profile symmetric in the $x$ direction:
\begin{equation}
\label{ns} n_s=\frac{1}{6\pi^2}\left(\frac{2m}{\xi\hbar^2}\right)^{3/2}\left(\mu_0-\frac{1}{2}m\omega_\perp^2 r_\perp^2+\frac{1}{2}m\omega_x^2(x^2+d^2)\right)^{3/2}.
\end{equation}
The potentials trapping the spin-up and spin-down components have instead minima at $x=\pm d$ respectively, thereby favouring the formation of a  spin polarized configuration with density
\be
\label{np}
n_{\ua\da}^p=\frac{1}{6\pi^2}\left(\frac{2m}{\hbar^2}\right)^{3/2}\left(\mu_0-\frac{1}{2}m\omega_\perp^2r_\perp^2-\frac{1}{2}m\omega_x^2(x\mp d)^2)\right)^{3/2}.
\ee

\begin{figure}[ptb]
\begin{center}
\includegraphics[height=4.5cm] {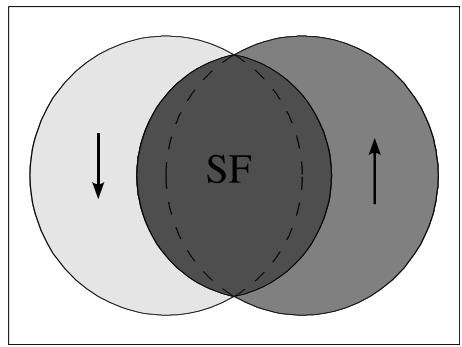}
\end{center}
\caption{Sketch of a central slice of the dipolar configuration as given in Eq. (\ref{eq:boundary}).}
\label{fig:dipoleconf}
\end{figure}
The boundary separating the two density profiles is fixed by the condition (\ref{eqcond}) of mechanical equilibrium and is characterized by the typical geometry of Fig. \ref{fig:dipoleconf} 
where, for sake of simplicity, we have shown a 2D cut.
For example for $x>0$ we have $\mu_p=\mu_\ua$ and the boundary between the superfluid and the spin-up normal component is fixed by the condition 
\be \left(x+\alpha d\right)^2+\lambda r_\perp^2=R_x^2+\gamma d^2,\label{eq:boundary}\ee
where we have introduced the radius $R_x=2\mu_0/m\omega_x^2$ of  the polarized component,   $\lambda=\omega_\perp/\omega_x$, $\alpha=(2\xi)^{3/5}/(2-(2\xi)^{3/5})$ and $\gamma=4 ((2\xi)^{3/5}-1)/(2-(2\xi)^{3/5})^2$.
We find that, for any value $d<R_x^0$, where $R_x^0=a^{ho}_x(24N)^{1/6}$ is the Thomas-Fermi radius of the non-interacting cloud,  with $a^{ho}_x=\sqrt{\hbar/m\omega_x}$, there is equilibrium between the superfluid and the spin-polarized phases. Conversely, if $d>R_x^0$, the superfluid is absent and the two spin polarized clouds are separated in space. 

Starting from results (\ref{ns}) and (\ref{np}) it is possible to calculate the spin dipole moment (\ref{D})
as a function of the displacement $d$ of the traps.  
The spin-up (-down) densities entering the integral (\ref{D}) are given by the sum of the superfluid and spin polarized components: $n_{\ua\da}=n_s+n_{\ua\da}^p$. Since the superfluid density is $x$-symmetric, it does not contribute to the integral and the spin dipole moment turns out to be reduced  with respect to the value
$D=d$ predicted in the absence of interactions. 
\begin{figure}[ptb]
\begin{center}
\includegraphics[height=6cm] {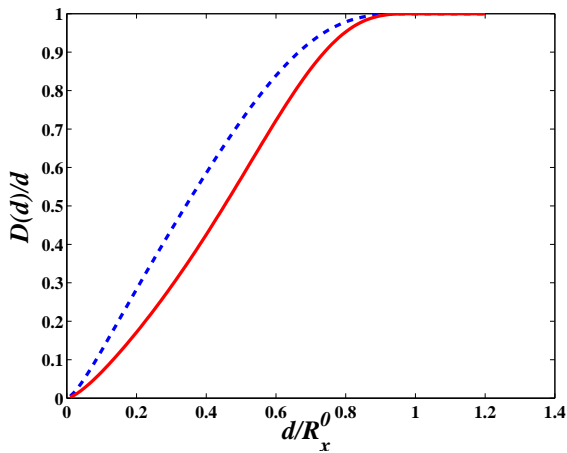}
\end{center}
\caption{Normalized induced spin dipole moment, $D(d)/d$, vs. displacement $d$ of the trapping potential. The displacement is given in units of the Thomas-Fermi radius $R_x^0$}
\label{fig:dipole}
\end{figure}
In particular, when $d\ll R_x^0$ the induced dipole moment takes the simple  power law behaviour
\begin{equation}
\label{dsmall} 
D(d)\to\frac{2^{10}}{45\pi}\left(\frac{d}{R_x^0}\right)^{5/2}
\frac{\xi^{9/8}}{(2-(2\xi)^{3/5})^{5/2}},
\end{equation}
revealing explicitly that the magnetic response $D(d)/d$ vanishes in the linear limit $d\to 0$. Figure \ref{fig:dipole} shows that, due to superfluidity, the induced dipole moment deviates significantly from the non-interacting value $D=d$ at distances of the order of the size of the atomic cloud. At the same time the figure,
as well as Eq.(\ref{dsmall}), shows that the induced dipole is always different from zero even for small displacements, revealing the absence of a true gap. These two fetaures characterize the behaviour of the trapped superfluid at unitarity. Actually, near the border of the cloud, the gap, being proportional to the Fermi energy, becomes smaller and smaller and even a tiny magnetic field can induce a finite, although small, magnetization. 

The polarization of the gas becomes more and more pronounced on the BCS side of the resonance. In fact in this case the gap  decreases exponentially when we reach the border and, as a consequence, we can more easily polarize the medium by breaking Cooper pairs. For a trapped gas in the deep BCS regime we expect that the induced spin dipole moment approaches the value $D=d$ at small displacement distances, which can be estimated using energy arguments as $d\sim R_x^0 \sqrt{\Delta/\epsilon_F}$, where $\Delta$ is an average value of the superfluid gap. 

On the BEC side of the resonance, where molecules are formed at small density, we instead 
expect a very different behaviour, which can be described using 2-body physics. In fact in this case the gap remains finite even near the surface where it is fixed by the binding energy  $\hbar^2/ma^2$ of the free molecules. The polarization will be zero until the value of the displacement is such that the molecular potential energy increase $\sim m\omega_x^2 d^2$ is of the order of the molecular binding energy. This corresponds to separation distances of the order of 
$d/R_x^0\simeq 1/k_Fa$. In the deep BEC regime, where $k_Fa\ll 1$, we therefore expect that the system will never be polarized unless the displacement is much larger than the size of the cloud \cite{adiabaticity}. 

In conclusion, the transition from the BCS to the BEC regime is predicted to reveal dramatic behaviour of the induced dipole moment which  exhibits a transition from a quasi ideal regime where the gas is easily polarized and is only affected by superfluidity at small displacements (deep BCS), to a regime where the rigidity of molecules is strong and the system is not easily polarized by the separation of the trapping potentials (deep BEC).

In Fig. \ref{fig:dipoleconf} we have also shown the predictions for the induced spin dipole moment calculated using the phase diagram
of  the BCS approach  which is known to provide a semi-quantitive description of the equation of state at unitarity. In our approach the differences are due to the different value for the interaction parameter $\xi$ whose value is  $0.58$ instead of the correct value $0.44$. The different curves reveal the sensitivity of the spin-dipole curve to the proper description of the equation of state. They are easily interpreted since the smaller is the value of $\xi$, the more favorable is the superfluid phase and, consequently, the smaller is the polarizability. The BCS approach predicts that the phase diagram of figure \ref{fig:phases} includes a further, normal-mixed phase  \cite{erich} which is simply a noninteracting gas of spin-up and spin-down components. At unitarity we then predict a transition from a superfluid to a mixed-normal phase and from a mixed-normal to a spin polarized phase as we move from the center to the border of the cloud. The inclusion of this additional phase, within BCS model, does not result in any visible effect in the physical quantity $D(d)$.

Separating the two trapping potentials and inducing a spin dipole moment is not the only way to polarize a $N_\ua=N_\da\equiv N/2$ system. Another procedure consists in modifying the trapping frequencies of the two spin species, i.e, by choosing trapping potentials of the form 
%\be V_{\uparrow\downarrow}({\bf r})=\frac{1}{2}m \omega_{\uparrow\downarrow}^2|{\bf r}|^2 \; , \ee 
\be V_{\sigma}({\bf r})=\frac{1}{2}m (\omega_{\sigma,\perp}^2 r_\perp^2+\omega_{\sigma,x}^2 x^2) \; , \; \sigma=\ua,\da \; , \ee 
with, in general $\omega_{\ua,i}\ne\omega_{\da,i}$ for $i=x,\perp$. This gives rise to a relative compression of  the two spin clouds.   
In the following we consider the simplest case of isotropic trapping, for which the above procedure corresponds to inducing a spin monopole polarization where the radii of the two species take a different value. For a noninteracting gas we find $R_\ua^0-R_\da^0=(a^{ho}_\ua-a^{ho}_\da)(24N)^{1/6}$. In the presence of interactions the behaviour  is  quite different. In particular, even for large differences between the trapping frequencies the system remains fully superfluid and does not exhibit spin polarization unless the value of the interaction parameter $\xi$ is larger than $0.5$. This behaviour is well understood by a simple energetic argument. If the trapping frequency of one of the two spin species, say the spin-up component, is much smaller than the one of the other component, then the configuration with spin polarization would consist of two noninteracting, practically non overlapping  clouds with very different radii. The energy of this  configuration would be $E^p/N=(3/4)\hbar\omega_\da (3N)^{1/3}$, the contribution
 of the spin-up component being negligible. Conversely, the superfluid, feeling a confining potential with 
 $\omega_s^2=(\omega_\ua^2+\omega^2_\da)/2\simeq \omega^2_\da/2$, has an energy equal to $E^s=E^p\sqrt{2\xi}$. This  is smaller than $E^p$ since $\xi=0.44$. In other words, at least within the ``two phase description'', the system always prefers to form pairs and remain superfluid rather than giving rise to a phase sepration as happens in the dipole case. The effect of the asymmetric trapping is a change in the radius of the atomic cloud and in the collective oscillations as now the superfluid   feels a different effective trapping frequency.   Notice that this peculiar behaviour is typical of the unitary regime and of the BEC side of the resonance. On the BCS side one would expect a partial polarization of the system, since the energetic gain associated with the superfluid phase is smaller than at unitarity.

The analysis of the dipole (and monopole) configuration has been here carried out at unitarity within the easiest assumption of phase separation between a totally polarized normal Fermi gas and a fermionic superfluid and with $N_\ua=N_\da$. The generalization of our approach to 
$N_\ua\neq N_\da$ configurations is straightforward. Furthermore the inclusion of additional phases in the diagram of fig.1, resulting from the availability of more sophisticated microscopic theories, as well as of surface tension effects could be naturally accounted for in the calculation of the spin polarizability.

Concerning the experimental feasibility of the measurement of the spin polarizability a {\sl conditio sine qua non} is the possibility of producing spin-dependent trapping potentials. This in principle is feasible by profiting of the different polarization of the electronic spin  
in the two hyperfine states of the Fermi gas and working with magnetic gradients and/or  polarized laser optical trapping (see 
\cite{carusotto} and reference therein) . However at unitarity -- where the most interesting features concerning the BCS-BEC crossover take place and the superfluid phase is more easily achieved -- the electronic spin polarization of two hyperfine states can be very similar, due to the large value of the magnetic field usually required to reach Feshbach resonances, making the experimental measurement of the spin polarizability a difficult task. This is indeed the case for $^6$Li, while the situation looks more promising for $^{40}$K, due to the smaller value of the magnetic field at resonance. Let us finally remark that the formalism developed in this work  can be easily generalized to Fermi mixtures of different atomic species (e.g.,$^6$Li and $^{40}$K). In this case the achievement of an independent tuning of the the two trapping  potentials would  be much easier.

The authors like to thank F. Chevy and L. P. Pitaevskii for very helpful discussions.


\begin{thebibliography}{99}

\bibitem{Rice} G. B. Partridge, W. Li, R. I. Kamar, Y. Liao, R. G. Hulet, Science {\bf 311}, 503 (2006)

\bibitem{MIT1}  M. W. Zwierlein, A. Schirotzek, C. H. Schunck, W. Ketterle, Science {\bf 311}, 492 (2006). 

\bibitem{MIT2} M. W. Zwierlein, C. H. Schunck, A. Schirotzek, W. Ketterle, cond-mat/0605258.

\bibitem{theory} J. Carlson and S. Reddy, Phys. Rev. Lett. {\bf 95}, 060401
(2005); P. Pieri and G. C. Strinati
Phys. Rev. Lett. {\bf 96}, 150404 (2006); D. E. Sheehy and L. Radzihovsky, {\sl idem} 060401 (2006); J. Kinnunen, L. M. Jensen, and P. T\"orm\"a, {\sl idem}, 110403 (2006), and cond-mat/0604424; M. Haque, H.T.C. Stoof, cond-mat/0601321; W. Yi and L.-M. Duan
Phys. Rev. A {\bf 73}, 031604 (2006) and cond-mat/0604558;  K. Machida, T. Mizushime and M. Ichioka, cond-mat/0604339; C.-H. Pao, S.-K. Yip, cond-mat/0604530; M. M. Parish, F. M. Marchetti, A. Lamacraft, B. D. Simons, cond-mat/0605744.

\bibitem{chevy} F. Chevy, Phys. Rev. Lett. {\bf 96}, 130401 (2006). 

\bibitem{erich} T. N. De Silva, E. J. Mueller, cond-mat/0601314 and cond-mat/0604638

\bibitem{astraxi} J. Carlson, S.-Y. Chang, V. R. Pandharipande, and K. E. Schmidt, Phys. Rev. Lett. {\bf 91}, 050401 (2003); G. E. Astrakharchik, J. Boronat, J. Casulleras, and and S. Giorgini, Phys. Rev. Lett. {\bf 93}, 200404 (2004).

\bibitem{adiabaticity} In the deep BEC, even for very large displacements of the potentials, the probability for a molecule to dissociate will be small. In fact the time scale for an adiabatic displacement process has to be larger than the inverse of the minimum gap between the ground and the first excited state. For a molecule in the deep BEC a simple WKB estimate gives a minimum gap exponentially small for $d\gg a_{\rm ho}$: $\delta\simeq\exp\left[-2\left(d/a_{\rm ho}\right)^2\right]$.
This result is confirmed by the exact solution of the 2-body problem (Zbignew Idziaszek, private communication). 

\bibitem{carusotto}  I. Carusotto, J. Phys. B: At. Mol. Opt. Phys. {\bf 39}, S211 (2006)



\end{thebibliography}
\end{document}